\documentclass{elsart}

\usepackage{graphicx}% Include figure files
\usepackage{amssymb}% extra math symbols

\newif\ifpreprint
\preprinttrue
\def\aver#1{<\!\!#1\!\!>}

\begin{document}

\journal{Physics Letters B}

\begin{frontmatter}

\date{July 28, 2008}
%\date{\today}
\title{
\ifpreprint
\rightline{\normalsize UASLP--IF--08--001}
\vspace{-0.3cm}
\rightline{\normalsize FERMILAB--Pub--08--084--E}
\vspace{0.2cm}
\fi
First Observation of the Cabibbo-suppressed Decays
{\mbox{\boldmath$\Xi_c^+ \to \Sigma^+ \pi^- \pi^+$}} and 
{\mbox{\boldmath$\Xi_c^+ \to \Sigma^- \pi^+ \pi^+$}}
and Measurement of their Branching Ratios}

% This file is to be included in a tex document
% written in elsart.  To include it, just type
% \include filename at the proper location
% file produced at Mon Apr  2 11:30:48 2007

The SELEX Collaboration
\author[SLP]{E.~V\'azquez-J\'auregui},
\author[SLP]{J.~Engelfried\corauthref{cor}},
\corauth[cor]{Corresponding author.}
\ead{jurgen@ifisica.uaslp.mx}
\author[Iowa]{U.~Akgun},
\author[PNPI]{G.~Alkhazov},
\author[SLP]{J.~Amaro-Reyes},
\author[PNPI]{A.G.~Atamantchouk\thanksref{tra}},
\author[Iowa]{A.S.~Ayan},
\author[ITEP]{M.Y.~Balatz\thanksref{tra}},
\author[SLP]{A.~Blanco-Covarrubias},
\author[PNPI]{N.F.~Bondar},
\author[Fermi]{P.S.~Cooper},
\author[Flint]{L.J.~Dauwe\thanksref{tra}},
% ...
\address[Fermi]{Fermi National Accelerator Laboratory, Batavia, IL 60510, U.S.A.}
% ...
\address[ITEP]{Institute of Theoretical and Experimental Physics, Moscow, Russia}
% ...
\address[PNPI]{Petersburg Nuclear Physics Institute, St.\ Petersburg, Russia}
% ...
\address[SLP]{Universidad Aut\'onoma de San Luis Potos\'{\i}, San Luis Potos\'{\i}, Mexico}
% ...
\address[Iowa]{University of Iowa, Iowa City, IA 52242, U.S.A.}
\address[Flint]{University of Michigan-Flint, Flint, MI 48502, U.S.A.}
% ...
\thanks[tra]{deceased}
\thanks[trb]{Present address: Advanced Mask Technology Center, Dresden, Germany}
\thanks[trc]{Present address: University of California at Irvine, Irvine, CA 92697, USA}
\thanks[trd]{Present address: Instituto de F\'{\i}sica da Universidade Estadual de Campinas, UNICAMP, SP, Brazil}
\thanks[tre]{Present address: Kafkas University, Kars, Turkey}
\thanks[trf]{Present address: Physik-Department, Technische Universit\"at M\"unchen, 85748 Garching, Germany}
\thanks[trg]{Present address: The Boston Consulting Group, M\"unchen, Germany}
\thanks[trh]{Present address: Lucent Technologies, Naperville, IL}
\thanks[tri]{Present address: Baxter Healthcare, Round Lake IL}
\thanks[trj]{Present address: University of Alabama at Birmingham, Birmingham, AL 35294}
\thanks[trk]{Present address: NRCN, 84190 Beer-Sheva, Israel}
\thanks[trl]{Present address: S\"uleyman Demirel Universitesi, Isparta, Turkey}
\thanks[trm]{Present address: Legal Department, Oracle Corporation, Redwood Shores, California}
\thanks[trn]{Present address: DOE, Germantown, MD}
\thanks[tro]{Present address: Solidum, Ottawa, Ontario, Canada}
\thanks[trp]{Present address: Siemens Healthcare, Erlangen, Germany}
\thanks[trq]{Present address: Allianz Insurance Group IT, M\"unchen, Germany}

\begin{abstract}
We report the first observation of two Cabibbo--suppressed 
decay modes, 
$\Xi_c^+ \to \Sigma^+ \pi^- \pi^+$ and
$\Xi_c^+ \to \Sigma^- \pi^+ \pi^+$.
We observe $59\pm14$ over a background of $87$,
and $22\pm8$ over a background of $13$ events, respectively,
for the signals.
The data were accumulated using the SELEX spectrometer during the 1996--1997 
fixed target run at Fermilab, chiefly from a $600\,\mbox{GeV}/c$
$\Sigma^{-}$ beam.  
The branching ratios of the decays relative to the Cabibbo--favored
$\Xi_c^+ \to \Xi^-\pi^+\pi^+$ are measured to be
$B(\Xi_c^+\to\Sigma^+\pi^-\pi^+) /B(\Xi_c^+\to \Xi^-\pi^+\pi^+) = 
     0.48 \pm 0.20$, and
$B(\Xi_c^+\to\Sigma^-\pi^+\pi^+) /B(\Xi_c^+\to \Xi^-\pi^+\pi^+) = 
     0.18 \pm 0.09$, respectively.
We also report branching ratios for the same decay modes of the
$\Lambda_c^+$ relative to $\Lambda_c^+\to pK^-\pi^+$.
\end{abstract}


\end{frontmatter}

\section{Introduction}
\begin{figure}[htb]
\begin{center}
\hfill
\includegraphics[width=0.24\textwidth]{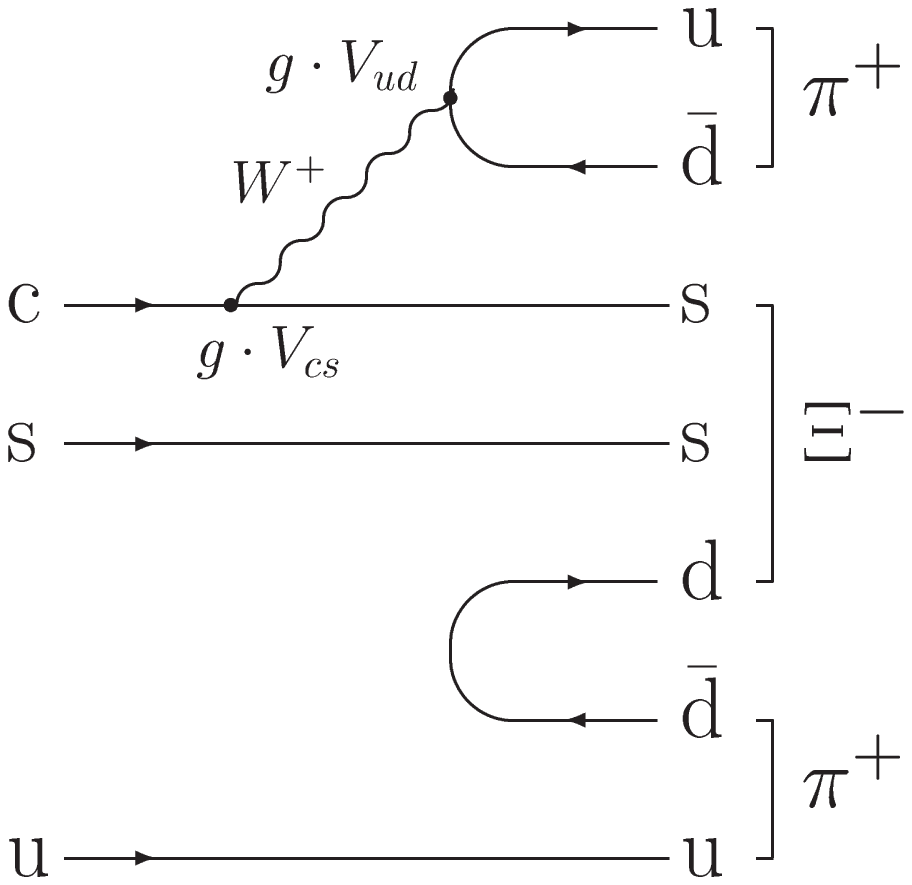}
\hfill
\includegraphics[width=0.24\textwidth]{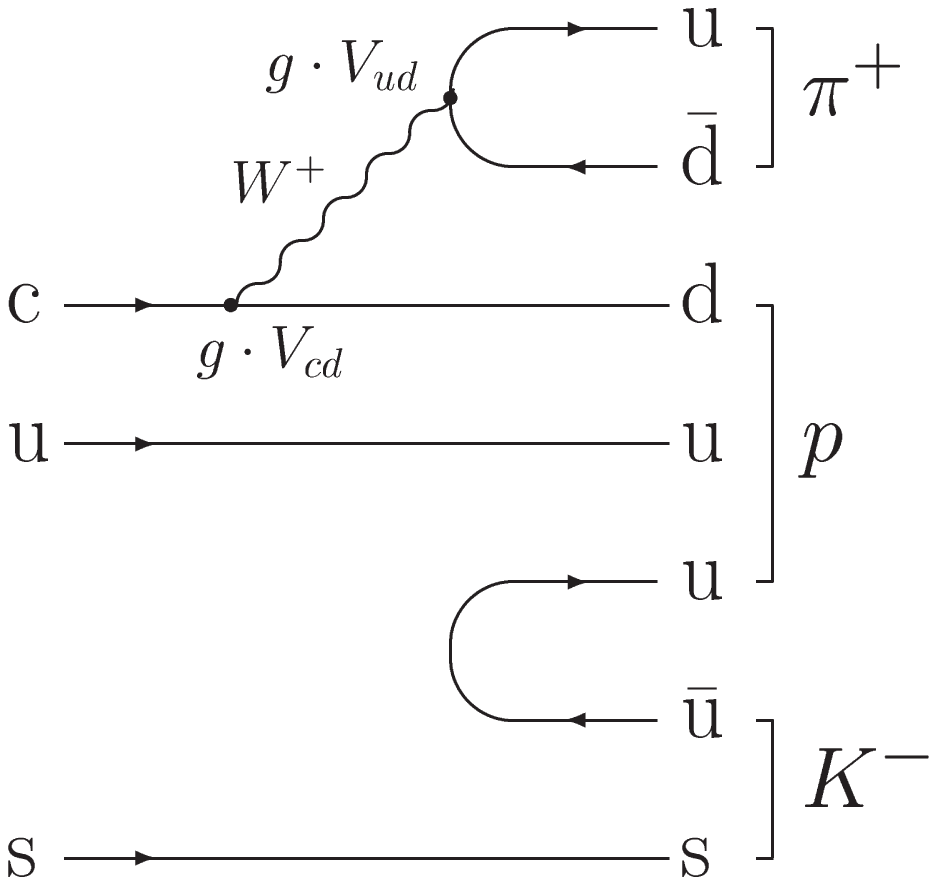}
\hfill
%\vspace{0.2cm}
%\hfill
\includegraphics[width=0.24\textwidth]{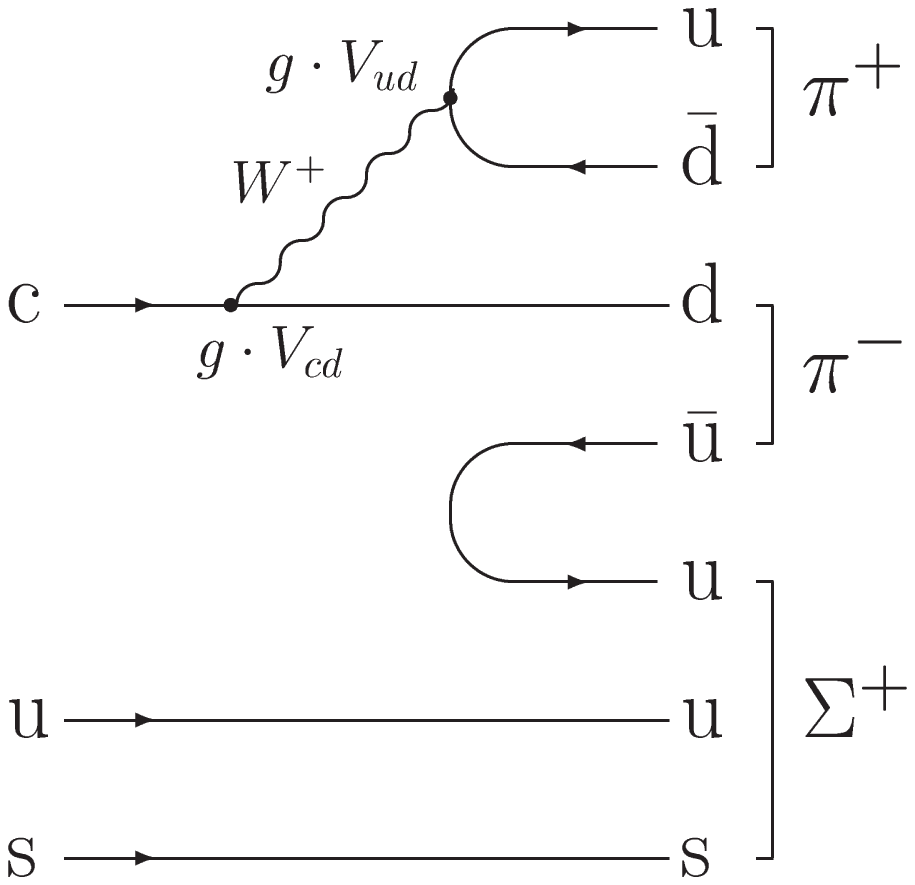}
\hfill
\includegraphics[width=0.24\textwidth]{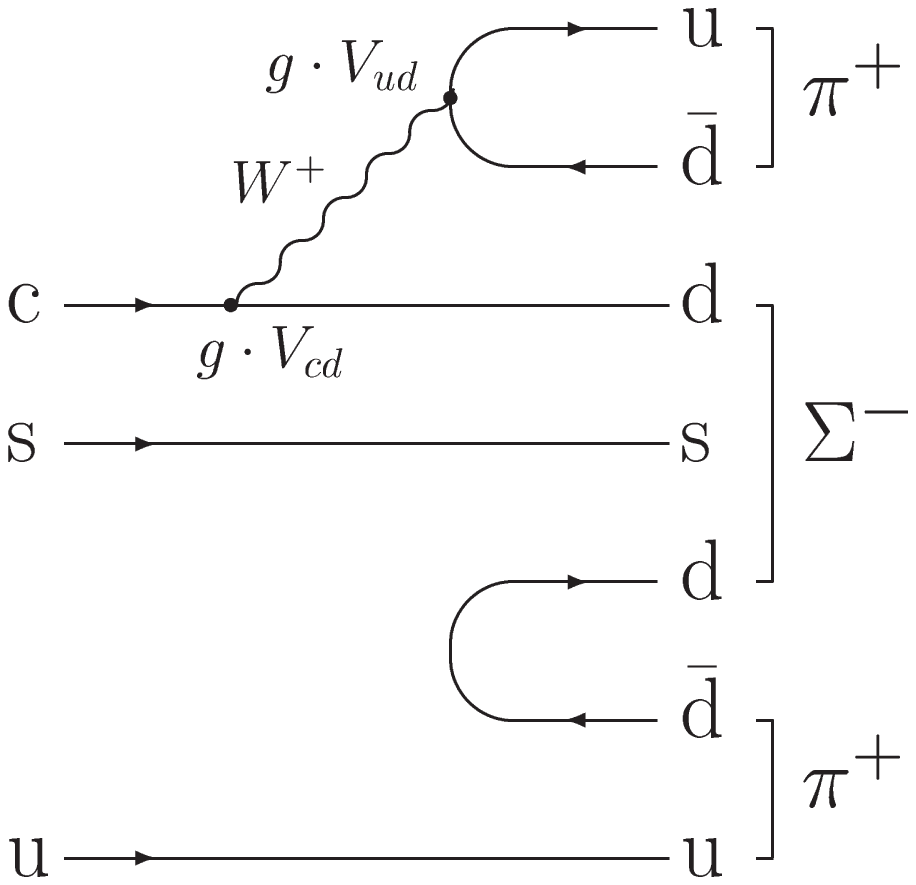}
\hfill{~}
\end{center}
\caption{Spectator diagrams for the decays (from left to right)
$\Xi_c^+\to\Xi^-\pi^+\pi^+$,
$\Xi_c^+\to pK^-\pi^+$,
$\Xi_c^+\to\Sigma^+\pi^-\pi^+$, and
$\Xi_c^+\to\Sigma^-\pi^+\pi^+$. 
The corresponding $W$-exchange diagrams and additional final-state quark
rearrangements are not shown here.}
\label{fig:specxc}
\end{figure}

\begin{figure}[htb]
\begin{center}
\hfill
\includegraphics[width=0.24\textwidth]{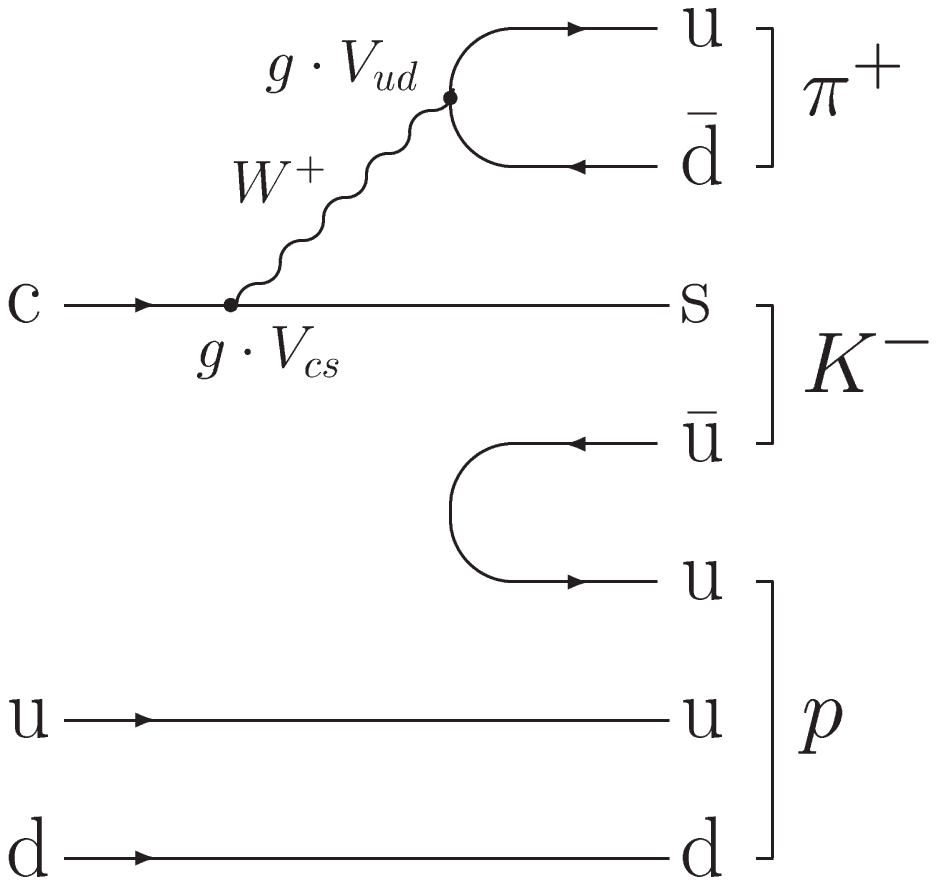}
\hfill
\includegraphics[width=0.24\textwidth]{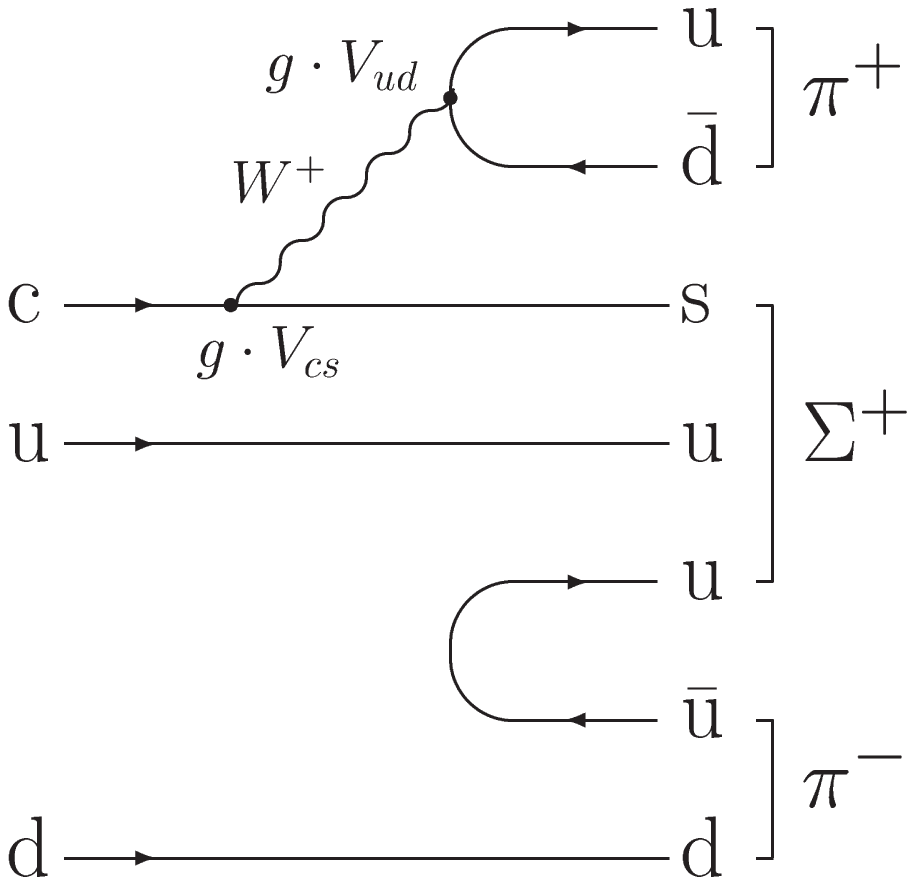}
\hfill
\includegraphics[width=0.24\textwidth]{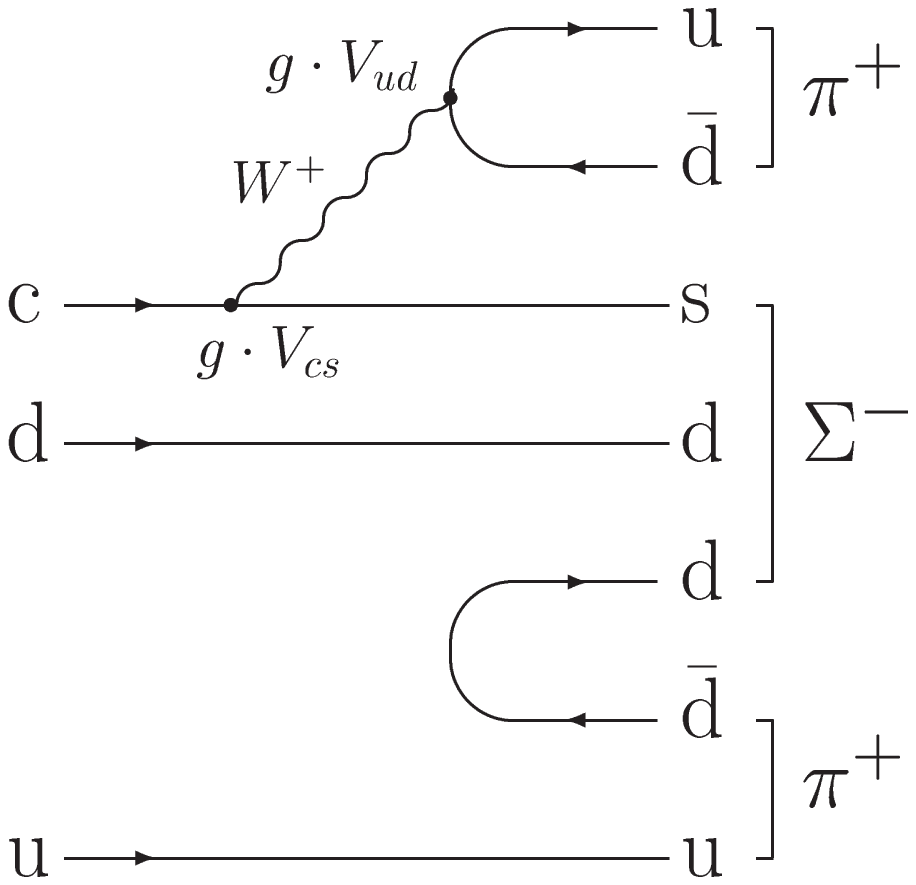}
\hfill{~}
\end{center}
\caption{Spectator diagrams for the decays 
$\Lambda_c^+\to pK^-\pi^+$ (left),
$\Lambda_c^+\to\Sigma^+\pi^-\pi^+$ (middle), and
$\Lambda_c^+\to\Sigma^-\pi^+\pi^+$ (right).
The corresponding $W$-exchange diagrams and additional final-state quark
rearrangements are not shown here.}
\label{fig:speclc}
\end{figure}

Studying Cabibbo--suppressed (CS) decays of hadrons provides insights 
into the weak interaction mechanism for non--leptonic
decays~\cite{Korner:1992wi}.
Comparing the strengths of CS decays to their
Cabibbo--favored (CF) analogs, one can, in a systematic way, assess the 
contributions of the various mechanisms.  In addition, comparing the same
or similar decay modes of different baryons allows some additional insights.
Even though any CS decay mode of the $\Xi_c^+$ is a CF mode of the
$\Lambda_c^+$, the detailed arrangement of the different final-state 
quarks into hadrons might be different, as shown in the spectator
diagrams in Fig.~\ref{fig:specxc} and Fig.~\ref{fig:speclc}.  While in the
case of the $\Lambda_c^+$ both final-state baryons, 
the $\Sigma^+$ and the $\Sigma^-$, can
have the $s$~quark resulting from the CF~$c$ decay, in the case of the
$\Xi_c^+$ decays with the identical final-state hadrons,
only the $\Sigma^-$ can be formed with
the CS~$c$ decay product.
By comparing several decay modes of different hadrons 
some information about the importances of 
direct quark emission at the
decay stage and from quark rearrangement due to final--state
scattering might be obtained.

Modern methods for calculating non--leptonic decay rates of the charm hadrons 
employ heavy quark effective theory and the factorization
approximation~\cite{Bauer:1986bm}.
Nonetheless, the three--body decays of charm 
baryons are prohibitively difficult to calculate due to the complexity of 
associated final-state interactions.  Measurements of the relative branching 
fractions of charm baryon states, both CF and CS, give additional information
about the structure of the decay amplitude and the validity of the 
factorization approximation.

Until now, the only CS $\Xi_c^+$ decays reported are
$\Xi_c^+\to pK^-\pi^+$~\cite{Jun:1999gn,Link:2001rn} and
$\Xi_c^+\to\Sigma^+K^-K^+$~\cite{Link:2003cd}.
In this letter, we present the first observations of 
$\Xi_c^+ \to \Sigma^+ \pi^- \pi^+$ and 
$\Xi_c^+ \to \Sigma^- \pi^+ \pi^+$, and  determine their branching ratios
relative to the CF
$\Xi_c^+\to\Xi^-\pi^+\pi^+$.
To validate our analysis method, we also report the branching ratios
$B(\Lambda_c^+\to\Sigma^+\pi^-\pi^+)/B(\Lambda_c^+\to pK^-\pi^+)$ and
$B(\Lambda_c^+\to\Sigma^-\pi^+\pi^+)/B(\Lambda_c^+\to\Sigma^+\pi^-\pi^+)$ 
and compare them to previously reported
results~\cite{Barlag:1992jz,Kubota:1993pw,Frabetti:1994kt}.

\section{Experiment}

SELEX is a high energy hadroproduction 
experiment using a 3--stage spectrometer designed for high acceptance for
forward ($x_F\gtrsim0.1$) interactions. 
The main goal of the experiment is the study of production and decay
properties of charm baryons. 
Particles in the negative
($600\,\mbox{GeV}/c$, $\simeq50\,\%$ $\Sigma^-$, $\simeq50\,\%$ $\pi^-$) 
and positive beam 
($540\,\mbox{GeV}/c$, $\simeq92\,\%$ $p$, $\simeq8\,\%$ $\pi^+$)
were tagged by a beam  transition radiation detector.  
The data were accumulated from a five-foil segmented
target (2~Cu, 3~C, each separated by $1.5\,\mbox{cm}$) with a total 
thickness of  $5\,\%$ of an interaction length for protons.
The spectrometer had silicon strip detectors to measure the beam and outgoing
tracks, giving precision primary and secondary vertex reconstruction.
Momenta of particles deflected by the analyzing magnets were 
measured by a system of proportional wire chambers (PWCs), drift chambers and
silicon strip detectors.  Momentum resolution for a typical
$100\,\mbox{GeV}/c$ track
was $\sigma_p/p \approx 0.5\,\%$. 
Charged particle identification was performed with a Ring
Imaging Cherenkov detector (RICH)~\cite{Engelfried:1998tv}, which 
distinguished $K^{\pm}$ from $\pi^{\pm}$ up to $165\,\mbox{GeV/}c$.
The proton 
identification efficiency was $>95\,\%$ above proton threshold
($\approx 90\,\mbox{GeV}/c$). 
For pions reaching the RICH detector,
the total mis-identification probability
due to all sources of confusion was $<4\,\%$.

Interactions were selected by a scintillator trigger.  The trigger for charm
required at least 4 charged tracks after the targets as indicated by an
interaction counter and at least 2 hits in a scintillator hodoscope after the
second analyzing magnet.  It accepted about 1/3 of all inelastic interactions.
Triggered events were further tested in an on--line computational filter 
based on downstream tracking and particle identification information.
The on--line filter selected events that had evidence of a secondary vertex
from tracks completely reconstructed using the forward PWC spectrometer and
the vertex silicon.  This filter reduced the 
data size by a factor of nearly 8 at a cost of about a factor of 2 in 
charm yield.
From a total of $15.2\cdot10^9$ interactions  
during the 1996--1997 fixed target run about $10^9$ events were
written to tape.
A more detailed description of the apparatus can be found
elsewhere~\cite{Jun:1999gn,Russ:1998rr}.

\section{Data Analysis}
In this analysis, secondary vertex reconstruction was attempted when 
the $\chi^2$ per degree of freedom for the fit of the ensemble
of charged tracks to a single primary vertex exceeded 4. 
All combinations of tracks
were formed for secondary vertices 
(in a first step with $\chi_{\rm sec}^2<9$, but
harder cut values were applied at later stages) and 
tested against a reconstruction table that specified 
selection criteria for each charm decay mode.
Secondary vertices which occurred inside the volume of a target
were rejected.
Common identification criteria for the different decay modes were:
proton and kaon candidate tracks were required to be identified 
by the RICH detector to be at least as likely as a pion; 
if a pion candidate track reached the RICH detector, we applied
as a loose requirement that it had to have 
a likelihood of at least 10\,\%, otherwise it was always accepted;
hyperon ($\Sigma^\pm$, $\Xi^-$) decays were identified by disappearance 
of a track in a
limited decay interval ($5 - 12\,\mbox{m}$ downstream from the target),
requiring that the candidate track had hits 
in the tracking detectors 
before the first and 
in-between the first and second magnet,
but no hits assigned along the extrapolated trajectory in the
14~chambers after the second analyzing magnet; this category of tracks gives 
unique $\Sigma^+$ identification but is ambiguous between
$\Sigma^-$ and $\Xi^-$.
Additional ambiguities in the mass assignments may arise 
due to
loose particle identification criteria for $p$, $K^\pm$, and
$\pi^\pm$; tighter cuts on the identification criteria would 
reduce the accessible
momentum range and the number of observed events.

%During the production phase of the data analysis candidate events were
%selected when the selection criteria were fulfilled and
%the invariant mass of the charm candidate was within a pre-defined window.
%Unfortunately, in the case of $\Sigma^+\pi^-\pi^+$
%and $\Sigma^-\pi^+\pi^+$,
%this mass window was defined for the investigation of the
%identical decay modes of the $\Lambda_c^+$,
%resulting in an artificial cutoff slightly above the $\Xi_c^+$ mass, at the
%same value as for the $pK^-\pi^+$ mode in~\cite{Jun:1999gn}.
 
As additional cuts with variable values depending on the decay modes
and the relative branching ratio to be determined we used:
\begin{itemize}
\item
the separation between the primary and secondary vertices in units
of its error ($L/\sigma$) and
the error itself ($\sigma$);
\item
the reconstructed charm momentum vector point-back to the primary
vertex, expressed as
the square of the 
distance of the reconstructed charm momentum vector to the primary vertex
in the target plane in units of its error ($pvtx$);
\item
the sum of the squares
of the transverse momenta of the daughter tracks with
respect to the charm hadron direction of flight ($\Sigma_{pt^2}$);
\item
the second-largest miss-distance of the daughter tracks in the target 
plane in units of its error ($scut$);
\item
minimum momenta for the
$\pi^\pm$ ($p_\pi$) and hyperon ($p_{\rm hyp}$) daughter tracks.
\end{itemize}
The selection criteria and the actual
values for these cuts are discussed in the following sections
and are listed in
tables~\ref{tab:xcnumbers} and~\ref{tab:lcnumbers}.

The total acceptance (geometrical acceptance and reconstruction efficiencies)
for the different decay modes of interest 
was estimated by embedding Monte Carlo charm decay
tracks into data events.  Momentum and energy were not conserved in the 
process, but studies indicate this has little effect on the single--charm
acceptance calculation.  Events were generated with an average transverse 
momentum $\aver{p_{T}} = 1.0\,\mbox{GeV}/c$ and longitudinal momentum 
distributions according to $(1-x_F)^n$, with $n=2.5$
($n=2.45\pm0.18$ for
$\Lambda_c^+$ production
with a $\Sigma^-$ beam~\cite{Garcia:2001xj}).
The value of $n$ was varied during
the systematic studies and did not affect the final branching ratio results.
Detector hits, including resolution
and multiple Coulomb scattering smearing effects, produced
by these embedded tracks were folded into the hit banks of the underlying data 
event.  The new ensemble of hits was passed through the SELEX off--line 
software.  The acceptance is the ratio of the number of reconstructed events
to the number of embedded events in a particular mode.  For the determination
of the branching ratios only the relative acceptances are relevant, leading 
to a cancellation of most systematic effects associated with the acceptance
corrections.

\section{First Observation of 
{\mbox{\boldmath$\Xi_c^+ \to \Sigma^+ \pi^- \pi^+$}} and
{\mbox{\boldmath$\Xi_c^+ \to \Sigma^- \pi^+ \pi^+$}}}
\label{sec:evidence}
\begin{figure}[htb]
\begin{center}
\includegraphics[width=0.49\textwidth]{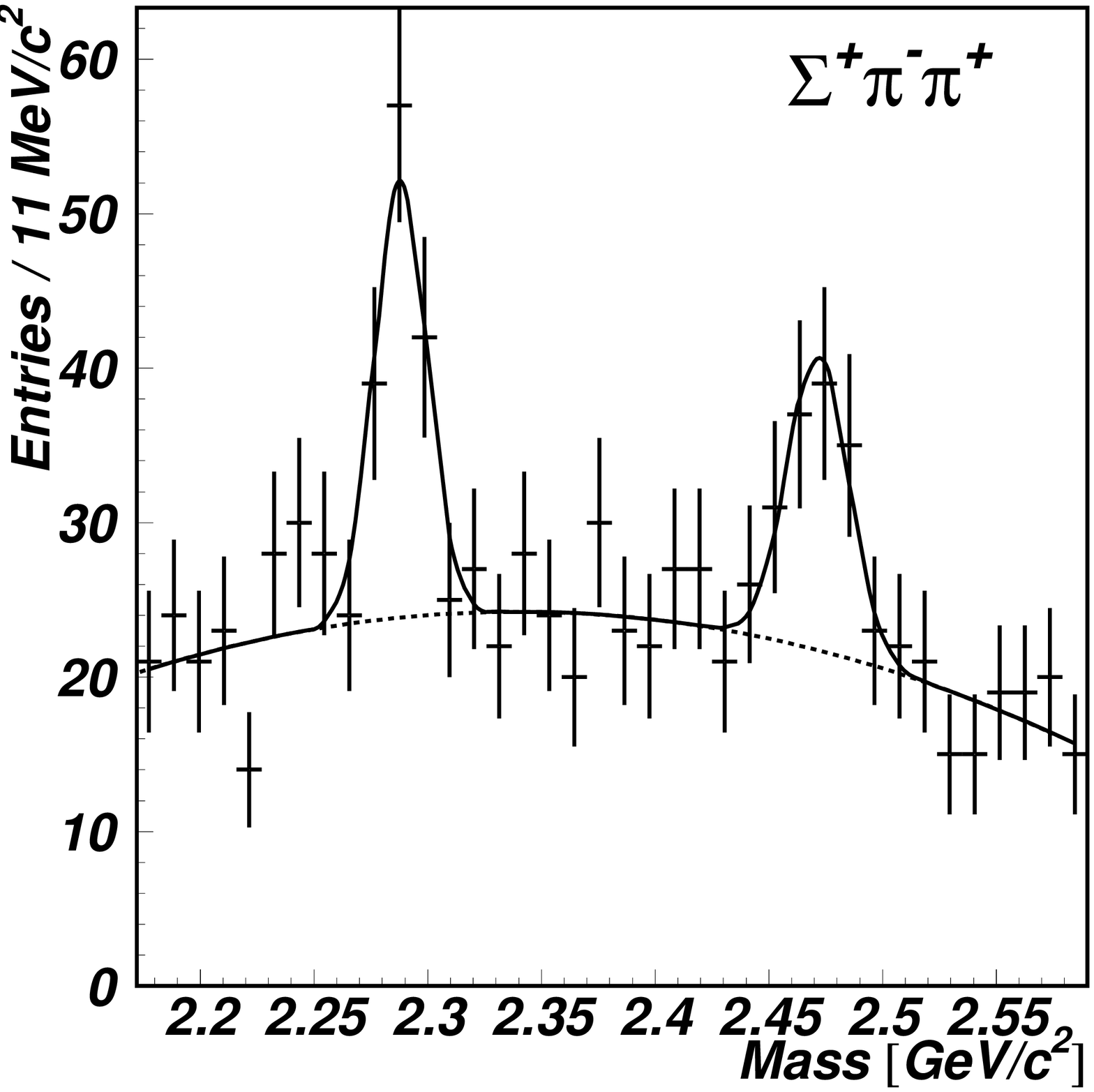}
\hfill
\includegraphics[width=0.49\textwidth]{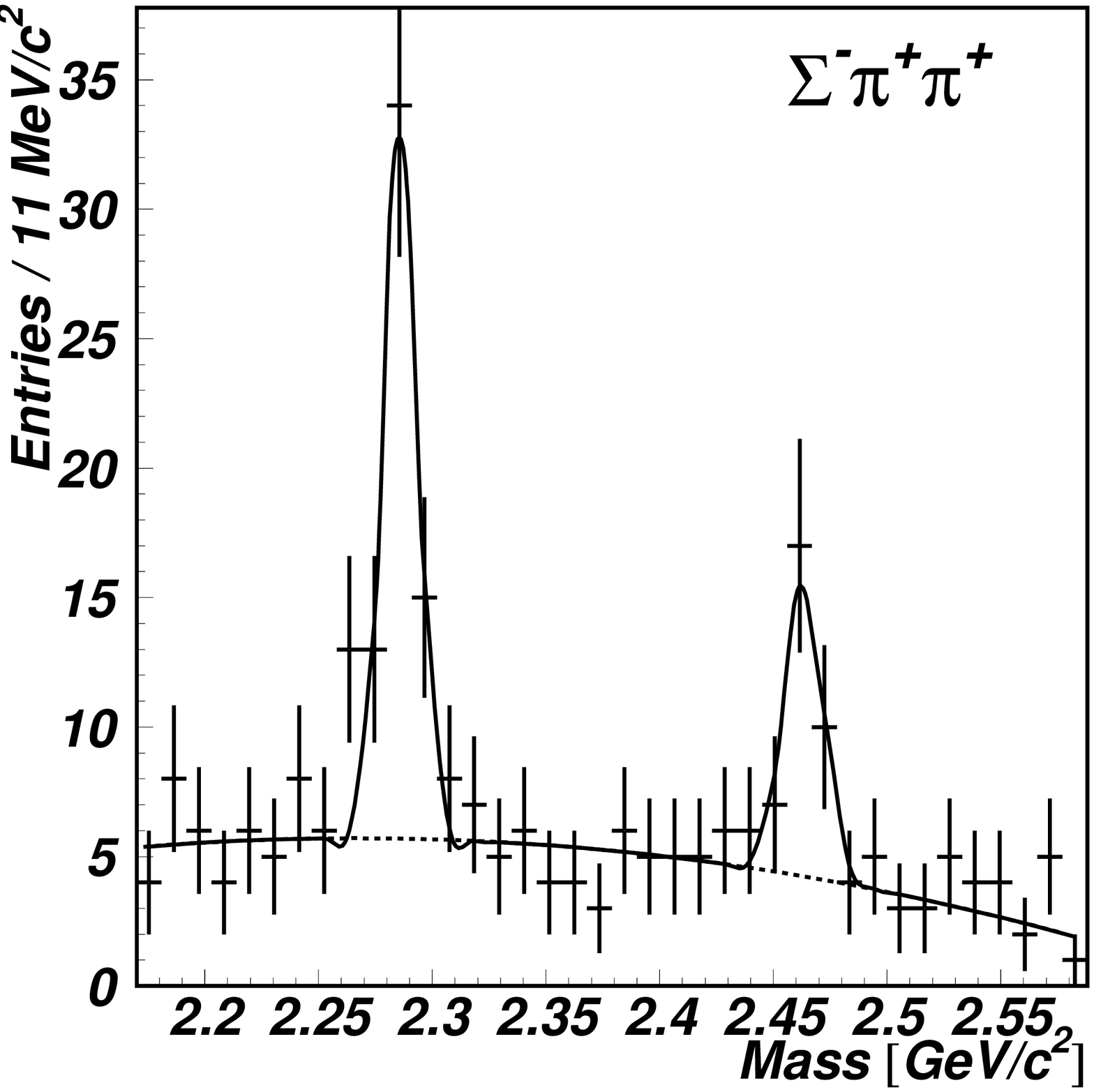}
\end{center}
\caption{Invariant mass distributions of $\Sigma^+\pi^-\pi^+$ (left)
and $\Sigma^-\pi^+\pi^+$ (right).}
\label{sii-evidence}
\end{figure}
In fig.~\ref{sii-evidence} we show the invariant mass distributions
of $\Sigma^+\pi^-\pi^+$ and $\Sigma^-\pi^+\pi^+$, over the full mass
range evaluated. 
%The bin widths are the experimental resolution
%for the corresponding $\Xi_c^+$ decays, as determined by Monte Carlo.
In each distribution we can see two peaks, corresponding to the $\Lambda_c^+$
and $\Xi_c^+$ decays.
The cuts used for the two distributions are shown in the first two
rows of tables~\ref{tab:xcnumbers} and~\ref{tab:lcnumbers}.
Additionally we required in both channels 
that at least one of the pions reached the
RICH detector.
For $\Sigma^+\pi^-\pi^+$ we applied
$\sigma<0.10\,\mbox{cm}$, $\chi_{\rm sec}^2<7$, $p_{\rm hyp}>70\,\mbox{GeV}/c$ and
events with an invariant mass around the $\Xi_c^+$ mass 
in the $\Sigma^+K^-\pi^+$ interpretation were removed;
for $\Sigma^-\pi^+\pi^+$ 
$\sigma<0.08\,\mbox{cm}$, $\chi_{\rm sec}^2<6$, $p_{\rm hyp}>80\,\mbox{GeV}/c$,
$scut>4$ and
events with an invariant mass around the $\Xi_c^+$ mass 
in the $\Xi^-\pi^+\pi^+$ interpretation were removed.
The selection of the individual cut values was based on prior SELEX analyzes
and tuned to suppress backgrounds mostly present in the
$\Xi_c^+$ mass region.
\begin{table}[htb]
\centering
\caption{Results of the Gaussian parts 
of the fits to the distributions presented in
fig.~\ref{sii-evidence}.}
\label{tab:evnumbers}
\begin{tabular}{|l|c|c||c|c|}
\hline\hline
\multicolumn{1}{|c|}{Mode} & 
Mass [$\mbox{MeV}/c^2$] &
Events &
Mass [$\mbox{MeV}/c^2$] &
 Events \\
\hline\hline
$\Sigma^+\pi^-\pi^+$ & 
$2288.1\pm2.2$ & $74.2\pm13.8$   & 
$2471.6\pm3.9$ & $58.7\pm13.5$ \\
\hline
$\Sigma^-\pi^+\pi^+$ & 
$2286.0\pm1.8$ & $46.4\pm10.1$   & 
$2463.3\pm3.0$ & $22.3\pm7.5$ \\
\hline
\end{tabular}
\end{table}
To both distributions we adjust the sum of two Gaussians 
with fixed widths (given by Monte Carlo) and a second degree polynomial.
The results of the fits are summarized in table~\ref{tab:evnumbers}.
For the decay $\Xi_c^+\to\Sigma^+\pi^-\pi^+$ we observe
$S=58.7\pm13.5$ signal events over a background of
$B=87.3\pm6.7$,
corresponding to a significance $S/\sqrt{B}=6.3\pm1.5$.
For the decay $\Xi_c^+\to\Sigma^-\pi^+\pi^+$ we observe
$S=22.3\pm7.5$ signal events over a background of $B=12.8\pm2.5$,
corresponding to a significance $S/\sqrt{B}=6.2\pm2.2$.
The masses of both the $\Lambda_c^+$ and the $\Xi_c^+$ are
slightly higher (in the case of $\Sigma^+\pi^-\pi^+$) and
lower ($\Sigma^-\pi^+\pi^+$) than the nominal values.
Varying the bin width and the fixed widths
of the Gaussians also gives consistent results.
We verified that the number of observed events varies as a function of
the cut variables, especially $L$ and $L/\sigma$, in the same
way as expected from Monte
Carlo (e.g.\ the observed events have the lifetime of the $\Xi_c^+$).

\section{Measurement of Branching Ratios}
\label{sec:br}

For the different branching ratio measurements we used 
different selection cuts,
chosen under
the criteria to minimize systematic effects on the final result.
We selected central cut values within a region where the Monte Carlo
described well the distributions of all the observables, for both the
decay mode of interest and the normalization mode.
If the same mode was used in different branching ratio determinations,
this selection could result in a different set of cuts.
We only used interactions initiated by a $\Sigma^-$ as beam particle and all
decay products (with the exception of the hyperons)
had to be within the RICH acceptance and correspondingly
identified. In the different modes we removed events stemming from 
the following reflections due to ambiguities in the mass assignments:
$D^+\to K^-\pi^+\pi^+$~(1),
$D^+\to K^+K^-\pi^+$~(2),
$D^+\to K^+\pi^-\pi^+$~(3),
$D_s^+\to K^+K^-\pi^+$~(4),
$D_s^+\to K^+\pi^-\pi^+$~(5),
$\Lambda_c^+\to p\pi^-\pi^+$~(6),
$\Lambda_c^+\to \Sigma^-\pi^+\pi^+$~(7),
$\Lambda_c^+\to pK^-\pi^+$~(8),
$\Xi_c^+\to\Xi^-\pi^+\pi^+$~(9),
$\Xi_c^+\to\Sigma^+K^-\pi^+$~(10).
We indicate the removed reflections and the corrections due to the removal
in tables~\ref{tab:xcnumbers} and~\ref{tab:lcnumbers}.

\begin{figure}[htbp]
\begin{center}
\includegraphics[width=\textwidth]{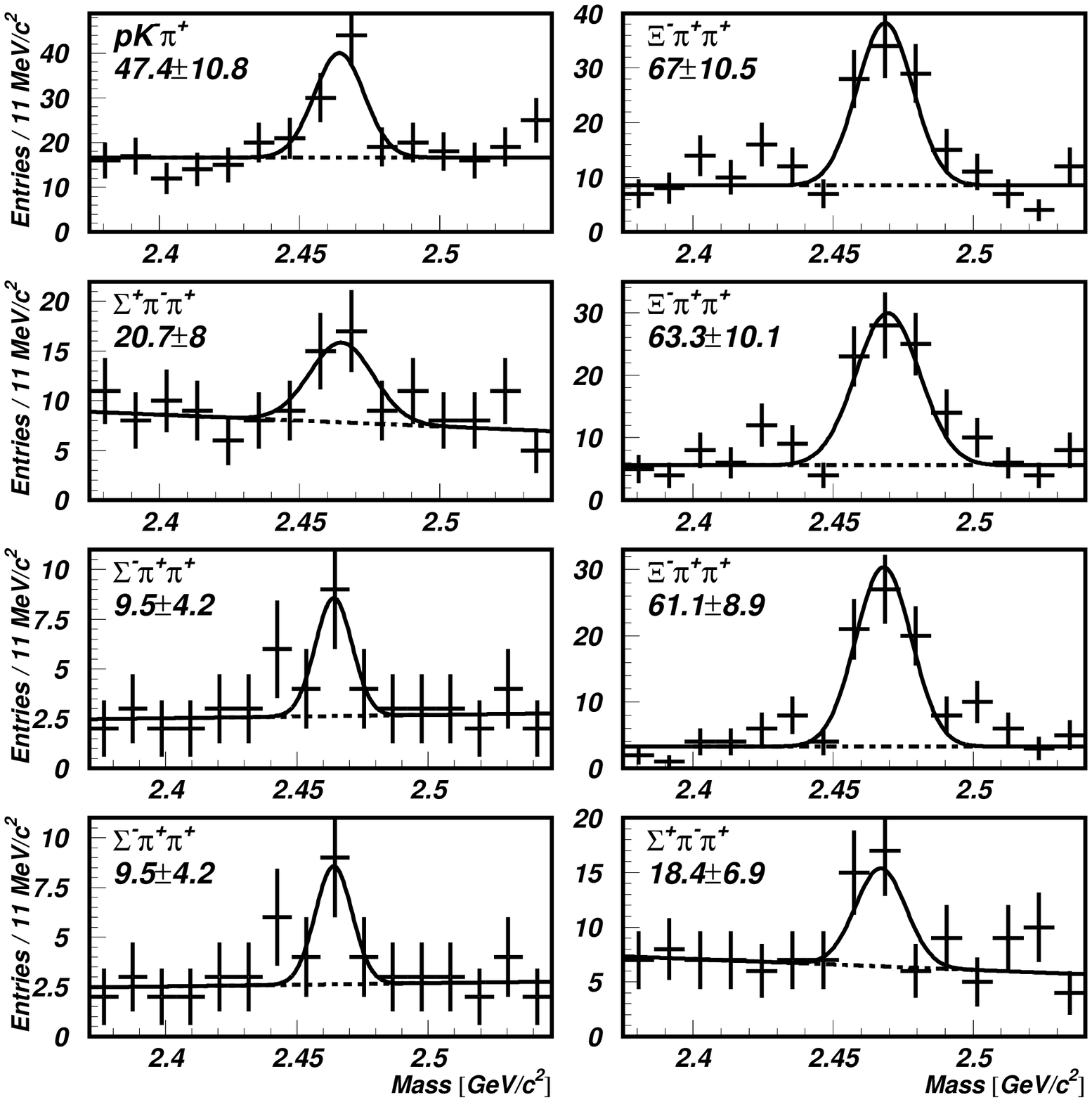}
\end{center}
\caption{Eight invariant mass distributions of:
$pK^-\pi^+$,
$\Sigma^+\pi^-\pi^+$, 
$\Sigma^-\pi^+\pi^+$, $\Xi^-\pi^+\pi^+$,
used to determine the four relative branching ratios
(in pairs from top to bottom)
$B(\Xi_c^+\to pK^-\pi^+) /B(\Xi_c^+\to \Xi^-\pi^+\pi^+)$,
$B(\Xi_c^+\to\Sigma^+\pi^-\pi^+) /B(\Xi_c^+\to \Xi^-\pi^+\pi^+)$,
$B(\Xi_c^+\to\Sigma^-\pi^+\pi^+) /B(\Xi_c^+\to \Xi^-\pi^+\pi^+)$,
and $B(\Xi_c^+\to\Sigma^-\pi^+\pi^+) /B(\Xi_c^+\to \Sigma^+\pi^-\pi^+)$,
respectively.
Different selection cuts were used for each branching ratio (see text).
We adjust a Gaussian (fixed width given by Monte Carlo) over a linear
background to each of the distributions.
The event yields are summarized in table~\ref{tab:xcnumbers}.}
\label{fig:brxc}
\end{figure}
In fig.~\ref{fig:brxc} 
we show the invariant mass distributions of 
$\Sigma^+\pi^-\pi^+$, $\Sigma^-\pi^+\pi^+$, $pK^-\pi^+$, and
$\Xi^-\pi^+\pi^+$ in the
$\Xi_c^+$ mass region.

\begin{figure}[htbp]
\begin{center}
\includegraphics[width=\textwidth]{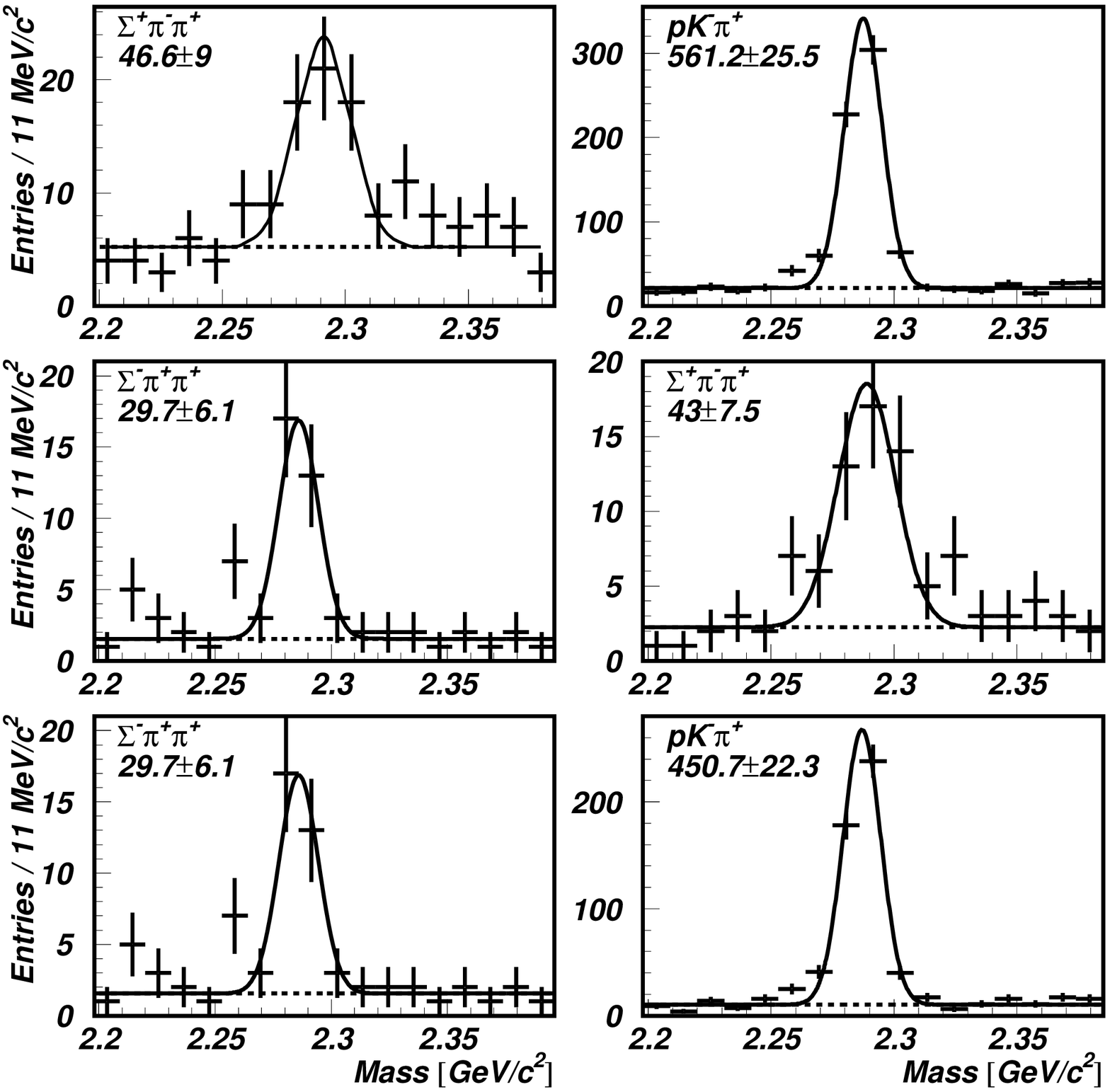}
\end{center}
\caption{Six invariant mass distributions of:
$\Sigma^+\pi^-\pi^+$, $pK^-\pi^+$,
$\Sigma^-\pi^+\pi^+$,
used to determine the three relative branching ratios
(in pairs from top to bottom)
$B(\Lambda_c^+\to\Sigma^+\pi^-\pi^+)/B(\Lambda_c^+\to pK^-\pi^+)$,
$B(\Lambda_c^+\to\Sigma^-\pi^+\pi^+)/B(\Lambda_c^+\to\Sigma^+\pi^-\pi^+)$,
and $B(\Lambda_c^+\to\Sigma^-\pi^+\pi^+)/B(\Lambda_c^+\to pK^-\pi^+)$,
respectively.
Different selection cuts were used for each branching ratio (see text).
We adjust a Gaussian (fixed width given by Monte Carlo) over a linear
background to each of the distributions.
The event yields are summarized in table~\ref{tab:lcnumbers}.}
\label{fig:br_lc}
\end{figure}

To verify our analysis method we determine the relative branching ratios
of two $\Lambda_c^+$ decay modes which are identical to our newly observed
$\Xi_c^+$ modes, using the fact that every CF~$\Lambda_c^+$ decay mode
is also a CS~$\Xi_c^+$ mode.  
In fig.~\ref{fig:br_lc} 
we show the invariant mass distributions of 
$\Sigma^+\pi^-\pi^+$, $\Sigma^-\pi^+\pi^+$, and $pK^-\pi^+$
in the $\Lambda_c^+$ mass region.

The number of observed events was determined by 
adjusting a Gaussian of fixed width (given by Monte Carlo) 
and a first-order polynomial to the distributions shown in
figs.~\ref{fig:brxc} and~\ref{fig:br_lc}\footnote{Counting the number of
entries above the extrapolated background, and using a second-order
polynomial for the background,
we obtain within errors the same number of events.}.
While most of the reflections lie outside
of the mass peaks and are removed to smoothen the backgrounds,  
some of them extend below the peaks and remove good events; we studied
this effect carefully with Monte Carlo
and corrected the number of observed events
for these losses. To determine the correction we simulated the shape of
the invariant mass distribution of the reflected mode (including all 
cuts) and scaled the number of events below the peak region to the number
of observed events in the reflected mode, keeping the same relative error
for the number of observed events.
We also studied correlations for the cases where more than one reflection
was removed and found them to be small and negligible.
The corrected yields for the different modes are,
along with the cuts used to obtain the distributions and the corresponding
total acceptances, presented in
tables~\ref{tab:xcnumbers} and~\ref{tab:lcnumbers}.
\begin{table}[htb]
\centering
\caption{Number of observed events and total acceptances for the different
$\Xi_c^+$ decay modes, with the corresponding cuts applied to each mode.
Common cuts are:
$\chi_{\rm sec}^2<8$,
$\sigma<0.10\,\mbox{cm}$, 
$scut>8$,
$p_{\rm hyp}>40\,\mbox{GeV}/c$.
The first two rows refer to the signals shown in fig.~\ref{sii-evidence},
with different common cuts as described in section~\ref{sec:evidence}.
``Corrected Events'' are
the number of observed events plus 
the corrections due to the removal of reflections and are shown 
separately in 
parenthesis; we keep the relative error from the fits. 
}
\label{tab:xcnumbers}
\begin{tabular}{|l|c|c|c|c|c|c|}
\hline\hline
\multicolumn{1}{|c|}{$\Xi_c^+$} & & &
$\Sigma_{pt^2}$&
Removed &
Corrected & Acceptance\\
\multicolumn{1}{|c|}{Mode} & 
$L/\sigma$ &
$pvtx$ &
[$\mbox{GeV}^2/c^2$]&
Reflections &
Events & [\%]\\
\hline\hline
$\Sigma^+\pi^-\pi^+$ & 
$>12$ & 
$<13$ &
 $>0.4$ &
 (10) &
$ 58.7\pm13.5$   & $0.750$ \\
\hline
$\Sigma^-\pi^+\pi^+$ & 
$>8$ & 
$<10$ &
$>0.5$ &
 (9) &
$22.3\pm7.5$ & $0.950$\\
\hline
$pK^-\pi^+$& & & & (1,2,4,6) & $(47.4+14.0)\pm14.0$    & $4.164$
\vspace{-0.4cm}\\
 & $>11$&$<13$&
$>0.3$& &
\vspace{-0.4cm}\\
$\Xi^-\pi^+\pi^+$& & & & (7) & $(67.0+2.4)\pm10.9$    & $0.915$\\
\hline
$\Sigma^+\pi^-\pi^+$ & & & & (6) &$(20.7+1.6)\pm8.6$   & $0.586$
\vspace{-0.4cm}\\
& $>13$&$<13$&
$>0.35$& &
\vspace{-0.4cm}\\
$\Xi^-\pi^+\pi^+$& & & & (7) & $(63.3+2.0)\pm10.4$   & $0.825$\\
\hline
$\Sigma^-\pi^+\pi^+$ & & & &(2,3,5,9)&$(9.5+5.0)\pm6.4$ & $0.988$
\vspace{-0.4cm}\\
 & $>13$&$<10$&
$>0.35$& &
\vspace{-0.4cm}\\
$\Xi^-\pi^+\pi^+$& & & & (7) &$(61.1+2.6)\pm9.3$   & $0.800$\\
\hline
$\Sigma^-\pi^+\pi^+$ & & & &(2,3,5,9)&$(9.5+5.0)\pm6.4$ & $0.988$
\vspace{-0.4cm}\\
 & $>13$&$<10$&
$>0.35$& &
\vspace{-0.4cm}\\
$\Sigma^+\pi^-\pi^+$& & & & (6) &$(18.4+1.5)\pm7.5$   & $0.570$\\
\hline
\end{tabular}
\end{table}

To obtain the branching ratios, we divided the number of 
observed (corrected) events of the two modes, and divided again by the
relative acceptance. The statistical error on the acceptance
is negligible, and most systematic errors cancel in the relative acceptance.

For the systematic studies we varied any single cut value, as
well as the parameter~$n$ for the $x_F$ distribution in the Monte Carlo
simulation, within some range and determined the branching ratio for every
set of parameters; for the set of cuts used we did not observe evidence of
any trend; all systematic variations
are small compared to the statistical error
and will be ignored in the final results 
since they would not affect the quadrature sum of the total error.

\begin{table}[htbp]
\centering
\caption{Number of observed events and total acceptances for the different
$\Lambda_c^+$ decay modes, with the corresponding cuts applied to each mode.
Common cuts are:
$\chi_{\rm sec}^2<4$,
$\sigma<0.10\,\mbox{cm}$, 
$scut>8$,
$p_{\rm hyp}>40\,\mbox{GeV}/c$.
The first two rows refer to the signals shown in fig.~\ref{sii-evidence},
with different common cuts as described in section~\ref{sec:evidence}.
``Corrected Events'' are
the number of observed events plus 
the corrections due to the removal of reflections and are shown 
separately in 
parenthesis; we keep the relative error from the fits. 
}
\label{tab:lcnumbers}
\begin{tabular}{|l|c|c|c|c|c|c|}
\hline\hline
\multicolumn{1}{|c|}{$\Lambda_c^+$} &  & &
$\Sigma_{pt^2}$&
Removed &
Corrected & Acceptance\\
\multicolumn{1}{|c|}{Mode} & 
$L/\sigma$ &
$pvtx$ &
[$\mbox{GeV}^2/c^2$]&
Reflections &
Events & [\%]\\
\hline \hline
$\Sigma^+\pi^-\pi^+$ & 
$>12$ & 
$<13$ & 
$>0.4$ &
 (10) &
$ 74.2\pm13.8$   & 0.450 \\
\hline
$\Sigma^-\pi^+\pi^+$ & 
$>8$ & 
$<10$ & 
$>0.5$ &
 (9) &
$46.4\pm10.1$ & 0.500\\
\hline
$\Sigma^+\pi^-\pi^+$& & & & (8) & $(46.6+3.0)\pm9.6$ & $0.292$
\vspace{-0.4cm}\\
 & $>11$& $<7$&
$>0.3$& & 
\vspace{-0.4cm}\\
$pK^-\pi^+$             & & & & -- & $(561.2+0.0)\pm25.5$ & $2.367$\\
\hline
$\Sigma^-\pi^+\pi^+$& & & & (5,9) & $(29.7+2.2)\pm6.6$ & $0.434$
\vspace{-0.4cm}\\
 & $>11$& $<4$&
$>0.4$& &
\vspace{-0.4cm}\\
$pK^-\pi^+$             & & & & -- & $(450.7+0.0)\pm22.3$ & $1.923$\\
\hline
$\Sigma^-\pi^+\pi^+$& & & & (5,9) & $(29.7+2.2)\pm6.6$ & $0.434$
\vspace{-0.4cm}\\
 & $>11$& $<4$&
$>0.4$& &
\vspace{-0.4cm}\\
$\Sigma^+\pi^-\pi^+$& & & & (8) & $(43.0+3.4)\pm8.1$ & $0.241$\\
\hline
\end{tabular}
\end{table}
The resulting branching ratios are shown, together with previously measured
values, in table~\ref{tab:results}.

\clearpage
\section{Discussion and Conclusions}
In table~\ref{tab:results} we summarize the results for the different
branching ratios measured in this work. Comparing our results with previously
measured ones (where available) shows good agreement. 
\begin{table}[htbp]
\centering
\caption{Results of the different Branching Ratios measured in this
analysis, and comparison to previously published results (if available).
Also shown is the $\alpha$-parameter (see text) 
for each branching ratio result.}
\label{tab:results}
\begin{tabular}{|c|c|c|}
\hline\hline
Branching Ratio&This Analysis&Other Measurements\\
\hline\hline
$B(\Xi_c^+\to\Sigma^+\pi^-\pi^+)~/$ &
 $0.48\pm0.20$ & -- \\
$B(\Xi_c^+\to \Xi^-\pi^+\pi^+)$ 
& $\alpha=6.4\pm2.7$ &
\\ \hline
$B(\Xi_c^+\to\Sigma^-\pi^+\pi^+)~/$ &
 $0.18\pm0.09$ & -- \\
$B(\Xi_c^+\to \Xi^-\pi^+\pi^+)$
 & $\alpha=2.5\pm1.2$ &
\\ \hline
$B(\Xi_c^+\to\Sigma^-\pi^+\pi^+)~/$ &
 $0.42\pm0.24$ & -- \\
$B(\Xi_c^+\to \Sigma^+\pi^-\pi^+)$
 & $\alpha=0.43\pm0.25$ &
\\ \hline
$B(\Xi_c^+\to pK^-\pi^+)~/$ &
$0.194\pm0.054$ &
    $0.234\pm0.047\pm0.022$ \cite{Link:2001rn} \\
~~~~$B(\Xi_c^+\to \Xi^-\pi^+\pi^+)$
& $\alpha=2.6\pm0.7$ & $0.20\pm0.04\pm0.02$ \cite{Jun:1999gn}
\\ \hline
$B(\Lambda_c^+\to\Sigma^-\pi^+\pi^+)~/$ &
$0.314\pm0.067$ & --\\
~~~~$B(\Lambda_c^+\to pK^-\pi^+)$
& $\alpha=0.30\pm0.07$ & \\
\hline
$B(\Lambda_c^+\to\Sigma^+\pi^-\pi^+)~/$&
$0.72\pm0.14$ 
& $0.74\pm0.07\pm0.09$ \cite{Kubota:1993pw}\\
~~~~$B(\Lambda_c^+\to pK^-\pi^+)$
& $\alpha=0.68\pm0.14$ & $0.54^{+0.18}_{-0.15}$ \cite{Barlag:1992jz}\\
\hline
$B(\Lambda_c^+\to\Sigma^-\pi^+\pi^+)~/$ &
$0.38\pm0.10$ 
& $0.53\pm0.15\pm0.07$ \cite{Frabetti:1994kt}\\
~~~~$B(\Lambda_c^+\to\Sigma^+\pi^-\pi^+)$
& $\alpha=0.39\pm0.11$ & 
\\ \hline
\hline
\end{tabular}
\end{table}

To quantify the effects of final-state quark rearrangements in the different
decays via the relevant relative matrix elements, we calculate
$\alpha$, which is defined as the measured relative
branching ratio corrected for phase space differences and, in the case
of comparing CF and CS modes, for the 
ratio of the CKM matrix elements 
($V_{cd}/V_{cs}=0.233\pm0.001$~\cite{Yao:2006px}).
We note that the $\alpha$-parameter
for 
$B(\Sigma^-\pi^+\pi^+)/B(\Sigma^+\pi^-\pi^+)$ 
is consistent
in the decays of both
the $\Lambda_c^+$ and the $\Xi_c^+$.
Comparing the
diagrams in fig.~\ref{fig:specxc} and~\ref{fig:speclc} we conclude
that the source of the 
final-state quark
does not affect
the relative matrix element significantly.

In summary, we observe for the first time the Cabibbo-suppressed decay
modes 
$\Xi_c^+\to \Sigma^+\pi^-\pi^+$ and $\Xi_c^+\to\Sigma^-\pi^+\pi^+$
and estimate their branching ratios.  With the same analysis method we also
analyze previously reported modes of both the $\Xi_c^+$ and the
$\Lambda_c^+$ and find good agreement.

\section*{Acknowledgment}
The authors are indebted to the staff of Fermi National Accelerator Laboratory
and for invaluable technical support from the staffs of collaborating
institutions.
This project was supported in part by Bundesministerium f\"ur Bildung, 
Wissenschaft, Forschung und Technologie, Consejo Nacional de 
Ciencia y Tecnolog\'{\i}a {\nobreak (CONACyT)},
Conselho Nacional de Desenvolvimento Cient\'{\i}fico e Tecnol\'ogico,
Fondo de Apoyo a la Investigaci\'on (UASLP),
Funda\c{c}\~ao de Amparo \`a Pesquisa do Estado de S\~ao Paulo (FAPESP),
the Israel Science Foundation founded by the Israel Academy of Sciences and 
Humanities, Istituto Nazionale di Fisica Nucleare (INFN),
the International Science Foundation (ISF),
the National Science Foundation (Phy \#9602178),
NATO (grant CR6.941058-1360/94),
the Russian Academy of Science,
the Russian Ministry of Science and Technology,
the Russian Foundation for Basic Research (grant 05-02-17869),  
the Secretar\'{\i}a de Educaci\'on P\'ublica (Mexico)
(grant number 2003-24-001-026),
the Turkish Scientific and Technological Research Board (T\"{U}B\.ITAK),
and the U.S.\ Department of Energy (DOE grant DE-FG02-91ER40664 and
DOE contract number DE-AC02-76CHO3000).


\begin{thebibliography}{99}

\bibitem{Korner:1992wi}
  J.~G.~K\"orner and M.~Kr\"amer,
  %``Exclusive nonleptonic charm baryon decays,''
  Z.\ Phys.\  C {\bf 55} (1992) 659.
  %%CITATION = ZEPYA,C55,659;%%

\bibitem{Bauer:1986bm}
  M.~Bauer, B.~Stech and M.~Wirbel,
  %``Exclusive Nonleptonic Decays of D, D(s), and B Mesons,''
  Z.\ Phys.\  C {\bf 34} (1987) 103.
  %%CITATION = ZEPYA,C34,103;%%

\bibitem{Jun:1999gn}
  S.~Y.~Jun {\it et al.}  [SELEX Collaboration],
  %``Observation of the Cabibbo suppressed decay Xi/c+ --> p K- pi+,''
  Phys.\ Rev.\ Lett.\  {\bf 84} (2000) 1857
  [arXiv:hep-ex/9907062].
  %%CITATION = PRLTA,84,1857;%%

\bibitem{Link:2001rn}
  J.~M.~Link {\it et al.}  [FOCUS Collaboration],
  %``Measurement of the relative branching ratio BR(Xi/c+ --> p+ K-
  %pi+)/BR(Xi/c+ --> Xi- pi+ pi+),''
  Phys.\ Lett.\  B {\bf 512} (2001) 277
  [arXiv:hep-ex/0102040].
  %%CITATION = PHLTA,B512,277;%%


\bibitem{Link:2003cd}
  J.~M.~Link {\it et al.}  [FOCUS Collaboration],
  %``Measurements of Xi/c+ branching ratios,''
  Phys.\ Lett.\  B {\bf 571} (2003) 139
  [arXiv:hep-ex/0305038].
  %%CITATION = PHLTA,B571,139;%%

\bibitem{Barlag:1992jz}
  S.~Barlag {\it et al.}  [ACCMOR Collaboration],
  %``An Observation Of Exclusive Lambda(C)+ Decays Into Sigma+ And Mesons,''
  Phys.\ Lett.\  B {\bf 283} (1992) 465.
  %%CITATION = PHLTA,B283,465;%%

\bibitem{Kubota:1993pw}
  Y.~Kubota {\it et al.}  [CLEO Collaboration],
  %``Measurement of exclusive Lambda(c) decays with a Sigma+ in the final
  %state,''
  Phys.\ Rev.\ Lett.\  {\bf 71} (1993) 3255.
  %%CITATION = PRLTA,71,3255;%%

\bibitem{Frabetti:1994kt}
  P.~L.~Frabetti {\it et al.}  [E687 Collaboration],
  %``First observation of the Sigma- pi+ pi+ decay mode of the Lambda(c) baryon
  %and its branching ratio relative to the Sigma+ pi+ pi- mode,''
  Phys.\ Lett.\  B {\bf 328} (1994) 193.
  %%CITATION = PHLTA,B328,193;%%


\bibitem{Engelfried:1998tv}
  J.~Engelfried {\it et al.},
  %``The SELEX phototube RICH detector,''
  Nucl.\ Instrum.\ Meth.\  A {\bf 431} (1999) 53
  [arXiv:hep-ex/9811001].
  %%CITATION = NUIMA,A431,53;%%

\bibitem{Russ:1998rr}
  J.~S.~Russ {\it et al.} [SELEX Collaboration], 
                  in \emph{ Proceedings of the
                 29th International Conference on High Energy Physics,}
                 1998, edited by A. Astbury \emph{et al.} (World Scientific,
                 Singapore, 1998) Vol.~II, 1259 [arXiv:hep-ex/9812031].
%  %%CITATION = HEP-EX/9812031;%%

\bibitem{Garcia:2001xj}
  F.~G.~Garcia {\it et al.}  [SELEX Collaboration],
  %``Hadronic production of Lambda/c from 600-GeV/c pi-, Sigma- and p  beams,''
  Phys.\ Lett.\  B {\bf 528} (2002) 49
  [arXiv:hep-ex/0109017].
  %%CITATION = PHLTA,B528,49;%%

\bibitem{Yao:2006px}
  W.~M.~Yao {\it et al.}  [Particle Data Group],
  %``Review of particle physics,''
  J.\ Phys.\ G {\bf 33} (2006), 1, c.f.\ page~142.
  %%CITATION = JPHGB,G33,1;%%

\end{thebibliography}
\end{document}